\documentclass[aps,prb,twocolumn,superscriptaddress]{revtex4}

\usepackage{graphicx}

\begin{document}

\title{Dilatometric study of
$\mathbf{Ni}_{2+x}\mathbf{Mn}_{1-x}\mathbf{Ga}$ under magnetic
field}

\author{A.~N.~Vasil'ev}
\affiliation{Low Temperature Physics Department, Moscow State
University, Moscow 119899, Russia}

\author{E.~I.~Estrin}
\affiliation{Institute of Physical Metallurgy, Central Research
Institute for Ferrous Metallurgy, Moscow 107005, Russia}

\author{V.~V.~Khovailo}
\affiliation{Institute of Fluid Science, Tohoku University, Sendai
980-8577, Japan}

\author{A.~D.~Bozhko}
\author{R.~A.~Ischuk}
\affiliation{Low Temperature Physics Department, Moscow State
University, Moscow 119899, Russia}

\author{M.~Matsumoto}
\affiliation{Institute of Multidisciplinary Research for Advanced
Materials, Tohoku University, Sendai 980-8577, Japan}

\author{T.~Takagi}
\author{J.~Tani}
\affiliation{Institute of Fluid Science, Tohoku University, Sendai
980-8577, Japan}

\begin{abstract}
A study of polycrystalline and single crystalline
Ni$_{2+x}$Mn$_{1-x}$Ga alloys by means of dilatometric and strain
gage techniques shows that large strain can be induced in the
temperature range of the martensitic transformation by application
of an external magnetic field. The biggest strain induced by
magnetic field was observed in the samples of chemical
compositions where the structural and magnetic transitions couple,
i.e. occur at the same temperature.
\end{abstract}

\maketitle

\section{Introduction}

Recently, much activity has been devoted to the development of
ferromagnetic shape memory alloys. Various intermetallic compounds
have been considered from the viewpoint of a combination of
ferromagnetic properties and structural transformations of
martensitic type. Among them are the body centered cubic or body
centered tetragonal structures of Fe-Ni-Cr, Fe-Ni-C and
Fe-Ni-Co-Ti [1-3], and face centered tetragonal structures of
Fe-Pt and Fe-Pd [4,5]. Since the magnetic energy is very small
compared to the energy stored in chemical bonds, the influence of
the magnetic field on the shape memory alloys can be seen only in
a limited interval near the temperature of martensite -- austenite
or austenite -- martensite transformation.

The above mentioned were ferrous shape memory alloys, however
there exist non-ferrous materials which also exhibit a well
pronounced shape memory effect in the ferromagnetic state. The
most intensively studied of them is a Heusler type alloy
Ni$_2$MnGa [6-9]. For the stoichiometric composition its Curie
temperature is $T_C = 376$~K and the temperature of austenite
cubic to martensite tetragonal transformation is $T_m = 202$~K
[10]. The partial substitution of Mn by Ni results in the decrease
of $T_C$ and in the increase of $T_m$ until these finally merge in
the range of compositions near Ni$_{2.19}$Mn$_{0.81}$Ga [11]. The
sample of this composition and of nearby compositions were the
subject of the present study.

\section{Results and discussion}

The measurements were performed on both polycrystalline and single
crystalline samples. The sample dimensions were typically $3\times
3\times 6$~mm. Studies on polycrystalline samples were done for
three compositions, namely Ni$_{2.16}$Mn$_{0.84}$Ga,
Ni$_{2.19}$Mn$_{0.81}$Ga and Ni$_{2.20}$Mn$_{0.80}$Ga. In the
Ni$_{2.16}$Mn$_{0.84}$Ga sample the Curie temperature was well
separated from the temperatures of structural transformation, so
that austenite -- martensite transition took place in a sample
being in a ferromagnetic state [11]. As shown in Fig.~1, the
structural transformation in this case was accompanied by a
significant increase in sample length, but it was virtually
non-sensitive to the external magnetic field.

\begin{figure}[h]
\begin{center}
\includegraphics[width=7cm]{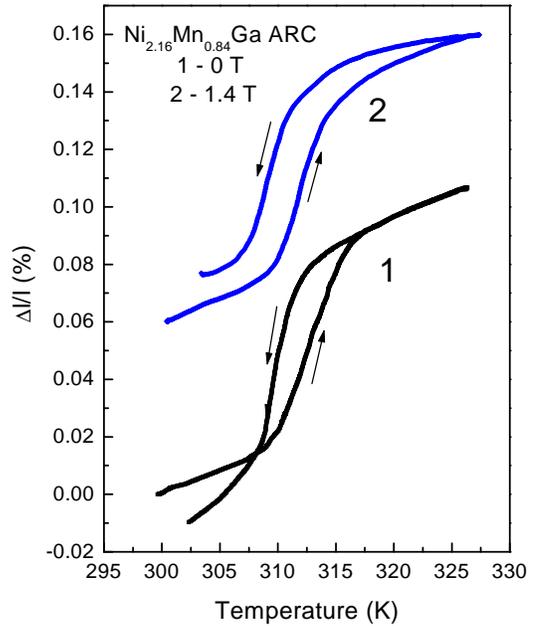}
\caption{Dilatometric effects in zero and 1.4~T magnetic fields
observed in the Ni$_{2.16}$Mn$_{0.84}$Ga polycrystalline sample.}
\end{center}
\end{figure}

In the Ni$_{2.19}$Mn$_{0.81}$Ga and Ni$_{2.20}$Mn$_{0.80}$Ga
samples a different picture at the phase transition was seen. For
these compositions the paramagnetic austenite transforms into
ferromagnetic martensite [11]. The transformation upon heating is
accompanied by a shortening of the sample. The change in the
sample length at phase transition was 0.04\% for
Ni$_{2.19}$Mn$_{0.81}$Ga and 0.12\% for Ni$_{2.20}$Mn$_{0.80}$Ga.
The same measurements performed in the presence of the 1.4~T
magnetic field oriented perpendicular to the long axis of the
sample showed a significant increase of the dilatometric effect of
transformation. In the Ni$_{2.19}$Mn$_{0.81}$Ga sample the effect
of transformation increases by 3.2 times, i.e. from 0.04\% to
0.13\%, in the Ni$_{2.20}$Mn$_{0.80}$Ga it increases by 2.6 times,
i.e. from 0.12\% to 0.31\%. The increase of the dilatometric
effect of transformation is seen also at cooling. The upward shift
of transition temperature due to the external 1.4~T magnetic field
is 1~K for Ni$_{2.19}$Mn$_{0.81}$Ga sample and 2.7~K for
Ni$_{2.20}$Mn$_{0.80}$Ga sample. The results of these measurements
are shown in Figs.~2 and~3.

\begin{figure}[t]
\begin{center}
\includegraphics[width=7cm]{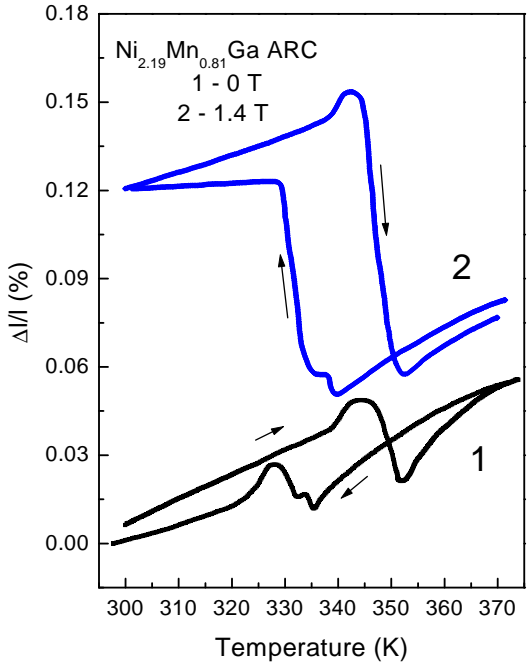}
\caption{Dilatometric effects in zero and 1.4~T magnetic fields
observed in the Ni$_{2.19}$Mn$_{0.81}$Ga polycrystalline sample.}
\end{center}
\end{figure}

\begin{figure}[h]
\begin{center}
\includegraphics[width=7cm]{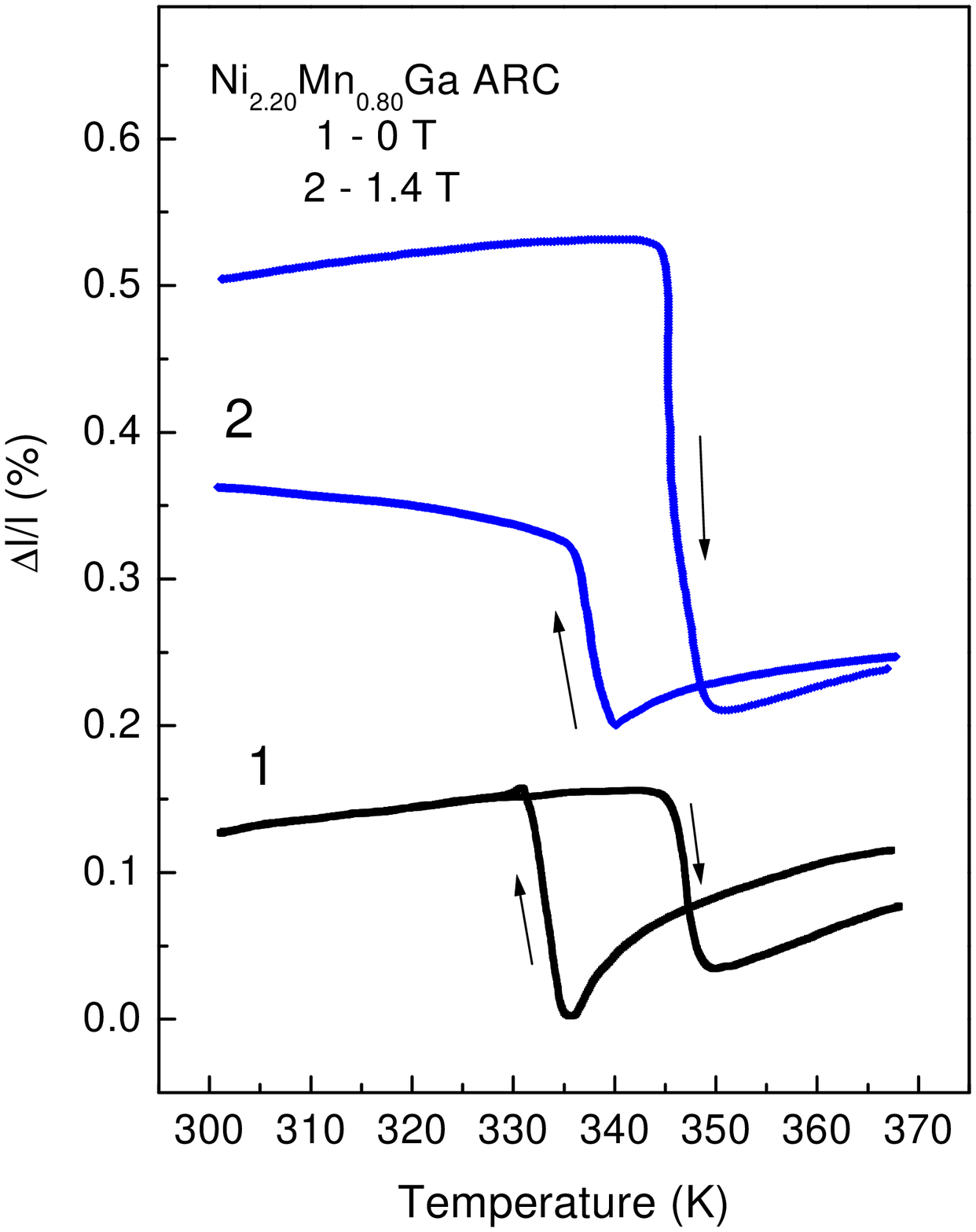}
\caption{Dilatometric effects in zero and 1.4~T magnetic fields
observed in the Ni$_{2.20}$Mn$_{0.80}$Ga polycrystalline sample.}
\end{center}
\end{figure}

\begin{figure}[h]
\begin{center}
\includegraphics[width=7cm]{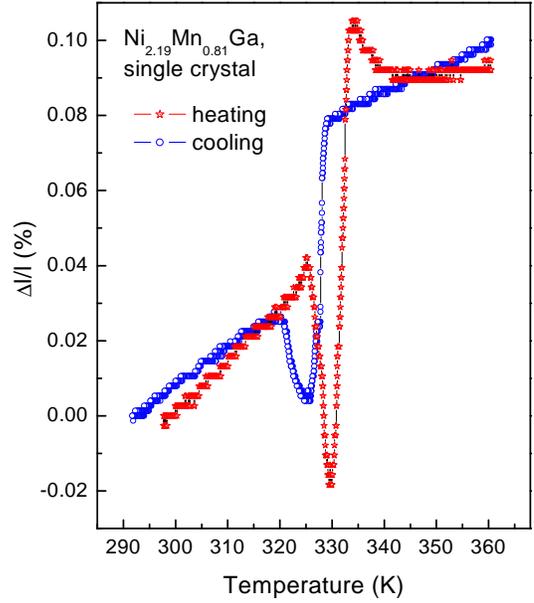}
\caption{Zero field strain in the Ni$_{2.19}$Mn$_{0.81}$Ga single
crystal upon heating and cooling.}
\end{center}
\end{figure}

The single crystal of Ni$_{2.19}$Mn$_{0.81}$Ga was grown from the
melt by the Czochralski technique. The dimensions of the as-grown
sample were about 80~mm in length and about 12~mm in diameter. No
special thermal treatment was employed to the crystal which showed
clear martensitic patterns on its surface at room temperature. The
crystal growth direction was [110]. The specimens for dilatometric
studies were spark-cut from the large single crystal and had
dimensions $3\times 3\times 6$~mm, where the longest dimension
coincided with the crystal growth direction. A non-magnetic strain
gage was attached along the crystal growth direction and the
sample was inserted into the variable temperature chamber of a
superconducting magnet. The phase transitions were easily
detectable by the sharp changes in strain gage response. Since the
phase transition in the sample studied is a first order it could
be characterized by $A_s$ and $A_f$, i.e. austenite start and
austenite finish temperatures at heating, which were found to be
325~K and 333.5~K, respectively. At cooling, the $M_s$ and $M_f$,
i.e. martensite start and martensite finish temperatures, were
found to be 329~K and 320.5~K, respectively. These data are
presented in Fig.~4. In the absence of a magnetic field the sample
exhibited elongation due to the thermal expansion in the
martensitic phase with the rate of about $12.5\times
10^{-6}/\mathrm{K}$. In the vicinity of phase transition it
exhibited a complicated behavior. Firstly, it shrinks by $-
6\times 10^{-4}$ in the range of 325 -- 330~K, then elongates by
$12\times 10^{-4}$ in the range of 330 -- 333.5~K and finally
exhibited elongation due to the thermal expansion in the
austenitic phase with the rate of about $5\times 10^{-6}$/K.

The application of a 5~T magnetic field along the largest
dimension of a sample resulted in the 5~K upward shift of the
critical temperatures. At 300~K the sample exhibited negative
magnetostriction of about $4.5\times 10^{-4}$ at saturating field.
At heating through the phase transition the sample shortens
initially by 0.09\%, then sharply elongates by 0.31\%, afterwards
it showed no temperature variation. At subsequent cooling, the
sample showed a step-like downward change of about 0.4\%. These
data are shown in Fig.~5.

\begin{figure}[t]
\begin{center}
\includegraphics[width=7cm]{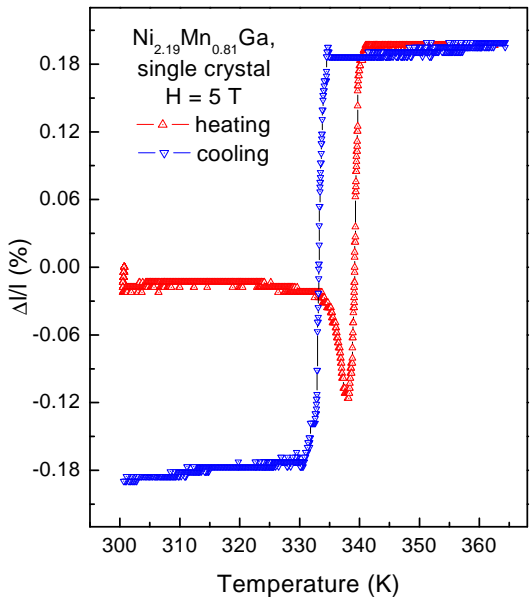}
\caption{Strain associated with the phase transition in
Ni$_{2.19}$Mn$_{0.81}$Ga single crystal in magnetic field 5~T.}
\end{center}
\end{figure}

\section{Conclusion}

In conclusion, it was shown that in the temperature interval of
the martensite -- austenite transformation the ferromagnetic
Ni$_{2+x}$Mn$_{1-x}$Ga alloys show the giant magnetostrictive
effects comparable with that observed in Terfenol-D. Taking into
account the narrowness of the transition temperature range these
observations open the possibility to induce complete phase
transformation by the application of a magnetic field and to reach
giant reversible magnetostrictive effect in the two-way trained
shape memory alloys.

\section*{Acknowledgements}

This work was partially supported by RFBR Grant No.~99-02-18247,
by the Grant-in-Aid for Scientific Research (C) No.~11695038 from
the Japan Society of the Promotion of Science and by the
"University of Russia" Program.

\end{document}